# Electrically Reconfigurable Non-Volatile On-Chip Bragg Filter with Multilevel Operation

Amged Alquliah, Jay Ke-Chieh Sun, Christopher Mekhiel, Chengkuan Gao, Guli Gulinihali, Yeshaiahu Fainman, and Abdoulaye Ndao[*]

Department of Electrical and Computer Engineering, University of California San Diego, La Jolla, California 92093, USA
[*]a1ndao@ucsd.edu

**Abstract:** Photonic integrated circuits (PICs) demand tailored spectral responses for various applications. On-chip Bragg filters offer a promising solution, yet their static nature hampers scalability. Current tunable filters rely on volatile switching mechanisms plagued by high static power consumption and thermal crosstalk. Here, we introduce, for the first time, a non-volatile, electrically programmable on-chip Bragg filter. This device incorporates a nanoscale layer of wide-bandgap phase change material ($Sb_2S_3$) atop a periodically structured silicon waveguide. The reversible phase transitions and drastic refractive index modulation of $Sb_2S_3$ enable dynamic spectral tuning via foundry-compatible microheaters. Our design surpasses traditional passive Bragg gratings and active volatile filters by offering electrically controlled, reconfigurable spectral responses in a non-volatile manner. The proposed filter achieves a peak reflectivity exceeding 99% and a high tuning range ($\Delta\lambda = 20$ nm) when transitioning between the amorphous and crystalline states of $Sb_2S_3$. Additionally, we demonstrate quasi-continuous spectral control of the filter stopband by modulating the amorphous/crystalline distribution within $Sb_2S_3$. Our approach offers substantial benefits for low-power, programmable PICs, thereby laying the groundwork for prospective applications in optical communications, optical interconnects, microwave photonics, optical signal processing, and adaptive multi-parameter sensing.

# 1. Introduction:

Programmable photonic integrated circuits provide a promising solution for scalable photonic architectures [1–4], enabling a range of novel on-chip functionalities such as AI accelerators [5], neuromorphic computing [6,7], quantum information processing [8,9], and microwave photonics [10–13]. Central to these circuits are programmable optical filters [14,7,15,1,10,12,16], which allow for real-time adjustment of the spectral response of guided light, facilitating applications like parallel optical computing [7,17], adaptive signal processing [12,14,18,19], astro-photonics [20,21], and multiparameter sensing applications [22–27]. Among the various PIC filters, integrated Bragg gratings stand out due to their excellent wavelength selectivity, high stability, broadband operation, and low insertion loss [14,28–32]. Unlike microring-based filters, integrated Bragg gratings are not constrained by free spectral range (FSR) limitations, which can hinder channel density [33–36]. Additionally, integrated Bragg gratings have improved side-mode suppression ratio, are robust against fabrication variation, and do not require complicated calibration of resonance conditions compared to microring-based filters [14,37–39].

However, the spectral response of conventional integrated Bragg gratings is fixed, limiting their tunability. Various thermo-optic and electro-optic methods have been explored to tune the spectral response [14,28,40–44]. However, these approaches suffer from high power consumption, limited tunability, large chip footprints, additional optical losses, considerable heat dissipation, and thermal crosstalk. A critical limitation of these techniques is their spectral response volatility, which requires a continuous high-power supply to maintain the desired spectral channel, making them impractical for low-frequency programmable photonic applications where infrequent tunability is needed [45,46], such as programmable microwave photonics, versatile photonic signal processing cores, integrated optical neural networks, and post-fabrication trimming.

Chalcogenide phase-change materials (PCMs) emerge as a powerful platform for a range of programmable photonic applications including optical memories [47], in-memory computing [48,49], neuromorphic computing [6,7], optical switching [50,51], metasurfaces [52–54], and post-fabrication trimming for fine-tuning device performance [45,55]. This versatility arises from the unique blend of properties in PCMs, such as reversible phase transitions between stable crystalline and amorphous states. Such transitions enable non-volatile operation, a critical attribute for maintaining programmed states without constant power consumption [56–58]. Moreover, PCMs are compatible with established CMOS fabrication processes, facilitating cost-effective and scalable photonic integration. Their threshold-switching behavior ensures excellent immunity to crosstalk, guaranteeing reliable operation in densely packed photonic environments [46,56–59]. The substantial refractive index modulation between the crystalline and amorphous phases facilitates the development of compact and efficient programmable photonic devices [60,61]. Additionally, PCMs offer wide spectral operability, and high endurance, capable of withstanding millions of switching cycles for reliable long-term performance [56,57,60,61].

Several methods can be employed to induce phase transitions in PCMs [60]. Thermal-conduction annealing shows irreversible switching, limiting device functionality to single-phase transition [62]. Optical excitation pulses provide high-speed switching but face challenges in generating and routing control signals [57]. Electrical switching offers a more practical solution due to its scalability and ability to achieve reversible and uniform switching of large PCM volumes [45,52–54,59,63,64]. This method enables low-energy operation and precise control over the amorphous/crystalline ratio within the material, facilitating numerous intermediate states. The significant advantages of electrical switching have driven recent research toward developing non-volatile, electrically programmable photonic devices. However, a non-volatile, electrically tunable on-chip Bragg filter has not yet been reported.

Here, we introduce the first nonvolatile, electrically programmable on-chip Bragg filter. By integrating a nanoscale layer of the wide-bandgap PCM ($Sb_2S_3$) on a chip-scale Bragg filter, we enable dynamic spatial tuning of the stopband of the filter using foundry-compatible microheaters. Our design surpasses traditional passive and active volatile filters with electrically controlled, reconfigurable spectral responses. The filter achieves over 99% peak reflectivity and a tuning range of ($\Delta\lambda = 20$ nm) between amorphous and crystalline states. We also demonstrate quasi-continuous spectral control (5-bit operation levels) by adjusting the amorphous/crystalline distribution within $Sb_2S_3$. This approach offers significant advantages for low-power, programmable PICs, enhancing applications in microwave photonics, optical signal processing, and integrated optical networks.

## 2. Filter design and working principle

**Figure 1A** schematically illustrates our proposed non-volatile, electrically programmable on-chip Bragg filter. The device comprises a nanoscale PCM layer patterned on the top surface of a distributed Bragg grating, formed using a silicon-on-insulator wafer with a thick $SiO_2$ top cladding. A metal nichrome microheater, positioned above the filter, enables uniform electrothermal switching of the PCM. The core principle of this device involves applying in-situ electrical pulses across the integrated metal microheaters, which raises the temperature of the PIC through Joule heating. This heat is transferred through the $SiO_2$ encapsulation layer into the $Sb_2S_3$ and Si waveguide beneath, inducing phase transitions between the amorphous and crystalline states of the $Sb_2S_3$, as well as intermediate states of $Sb_2S_3$. These phase transitions electrically reconfigure the entire index modulation profile of the grating, thereby dynamically tuning the device's spectral response. This capability offers a significant advantage over existing active Bragg gratings by enabling electrical and dynamic tuning in a nonvolatile, set-and-forget manner. Using foundry-compatible microheaters ensures seamless integration into existing photonic platforms, facilitating scalable photonic architectures.

The cross-sectional view **(Figure 1B)** details the layered structure of the device. It consists of a 50nm thick and 400nm wide $Sb_2S_3$ layer patterned atop the central surface of a sidewall corrugated silicon waveguide on an insulator. $Sb_2S_3$ was selected for its wide transparency window and high refractive index variation ($\Delta n$) [65]. The $Sb_2S_3$ film is encapsulated by a 50nm thick aluminum oxide ($Al_2O_3$) layer, which provides superior thermal conductivity and protection during annealing and thermal cycling. The addition of $Al_2O_3$ before the $SiO_2$ cladding is essential, as using only $SiO_2$ could impede the crystallization dynamics of the PCM [66]. The microheater comprises a resistive layer of nickel-chromium (NiCr), with dimensions of 200nm thickness and 1000nm width, connected to top gold (Au) contact pads via a 10nm titanium (Ti) adhesion layer. The heater is positioned on top of the SiO$_2$ cladding layer about 2μm away from the Bragg grating silicon waveguide to minimize optical insertion loss caused by the metal while enabling scalable control over the PCM states. NiCr was chosen for its high melting point and good thermal conductivity [67], which ensures the efficient heat transfer and electrical control necessary for the PCM.

The top view (**Figure 1C**) illustrates the structural parameters of the silicon Bragg grating waveguide. Periodic sidewall corrugations ($w$ and $\Delta w$) modulate the width of the Bragg grating, which is crucial for achieving the desired reflectivity and filtering characteristics. The grating period ($\Lambda$) is set to the target central wavelength of the stopband when the Sb$_2$S$_3$ is in its as-deposited amorphous state. The Bragg period is defined by the relationship $\Lambda_B = \frac{\lambda_B}{2\tilde{n}_{eff}}$ , where $\lambda_B$ is the Bragg wavelength and $\tilde{n}_{eff}$ is the average effective index of two widths of the Bragg structure. The device has a total length of $L = \Lambda \times N$, where $N$ is the number of grating periods.

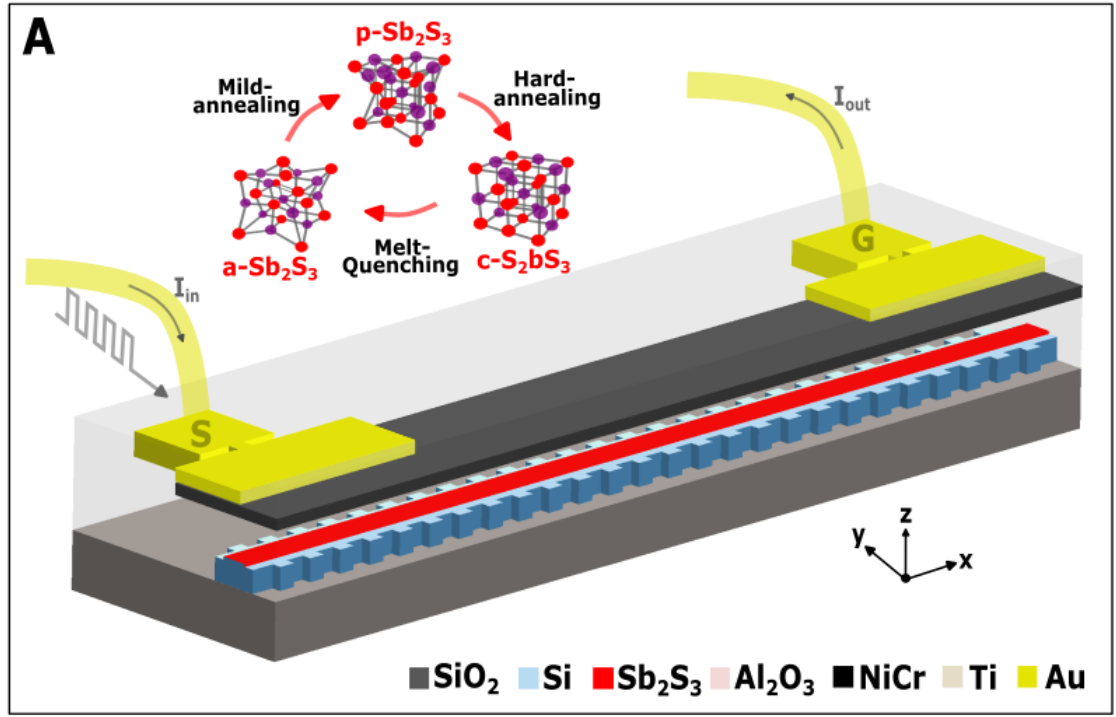

**Figure 1: Schematic of the Nonvolatile, Electrically Programmable On-Chip Bragg Filter.** (A) The device architecture comprises a periodically structured silicon nanophotonic waveguide with a superimposed nanoscale layer of wide-bandgap phase change material ($Sb_2S_3$) having a thickness of 50nm and a width of 400nm. The transition between the amorphous (a-$Sb_2S_3$), and crystalline (c-$Sb_2S_3$) states is achieved through annealing and melt-quenching processes. (B) Cross-sectional view showing the layered structure of the device, including the $Sb_2S_3$-clad integrated Bragg grating and microheater (C) Top view detailing the structural parameters of the filter.

## 3. Results and discussion

We investigated the distinct characteristics and responses of PCM switching in our programmable filter by performing mode analysis using the finite difference method (Ansys Lumerical) and heat transfer modeling using the finite element method (COMSOL). **Figure 2A-C** presents simulated electric field distributions of $TE_{00}$-like modes within the grating waveguide regions for both a-$Sb_2S_3$ and c-$Sb_2S_3$ states at $\lambda$=1581 nm. The waveguide's width varies between two sizes, denoted as *'w+Δw'* and *'w'*. The simulations were conducted at a wavelength of 1581 nm. The electric field distributions for the a-$Sb_2S_3$ cladding are shown in **Figures 2(a) and 2(c)**, while those for c-$Sb_2S_3$ are displayed in **Figures 2(b) and 2(d)**. Arrows overlaid on the electric field components indicate the polarization direction. The simulations show negligible insertion losses by the metallic microheater on top of the filter, highlighting the efficiency of our design. **Figure 2E** depicts the filter's tuning characteristics for both (a-Sb2S3) and (c-Sb2S3) states as a function of Bragg wavelength ($\lambda_B$). For a fixed $\Lambda_B$ = 320 nm, the Bragg condition is initially satisfied for a center wavelength $\lambda_B$ = 1581 nm when the $Sb_2S_3$ layer is in its amorphous state (point 1). Upon transitioning $Sb_2S_3$ to its crystalline state, the ($\tilde{n}_{eff}$) increases, as depicted in the inset Figure. This increase disrupts the original Bragg condition, suppressing the filtering response at $\lambda_B$ = 1581 nm. Consequently, the Bragg center wavelength undergoes a redshift, establishing a new Bragg condition at $\lambda_B$ = 1601 nm (point 2). **Supplementary Figure 1** further explores the relationship between the coupling coefficient ($\kappa$) (i.e., grating strength) and the $Sb_2S_3$ dimensions, elucidating the tunability mechanism.

Next, we examined the thermal response of the $Sb_2S_3$ by applying real-time voltage across the filter, as shown in **Figures 2F and 2G**. To simulate the phase transitions of the PCM during crystallization (Set) and amorphization (Reset), we employed COMSOL Multiphysics with a heat transfer model. These temperature simulations (**Figures 2C and 2E**) depict the thermal behavior of the PCM. The NiCr heater is positioned 2μm above the filter and extends 10μm beyond the grating to ensure uniform heating across the grating area. **Figures 2F and 2G** present the simulation results of the average temperature in the $Sb_2S_3$ patch overtime during the amorphization and recrystallization processes under a 4.5V, 10μs electrical pulse, and a 2V, 100μs electrical pulse, respectively. The insets show the corresponding cross-sectional temperature field at the peak temperature of each process. These results demonstrate the effective thermal management and precise control required for phase transitions in the $Sb_2S_3$ layer. Achieving and maintaining the necessary temperatures for amorphization and crystallization is crucial for the reliable performance of the programmable filter. **Supplementary Figure 2** provides a detailed analysis of the effective index, propagation loss, and thermal response of the proposed filter as a function of the heater position from the waveguide.

The switching speed of our programmable Bragg filter is influenced by the pulse width, trailing edge ($t_q$), and thermal relaxation time ($\tau$), which is defined as ($\frac{(T_m - T_g)}{t_q}$), where $T_m$, and $T_g$ are melting and glass transition temperatures, respectively [68]. During amorphization, which requires higher temperatures and shorter trailing edges, short programming pulses with trailing edges spanning several microseconds can be used. Conversely, crystallization requires lower temperatures and longer trailing edges, necessitating longer pulses with trailing edges in the tens of microseconds range. This switching speed aligns well with the requirements of our programmable photonic applications including our programmable filter [45,46,61]. The electrical pulse amplitudes and durations are carefully chosen to ensure the PCM exceeds its $T_m$ while surrounding materials remain below their respective melting points. The switching speed is not affected by the device length but is dependent on the kinetics of the PCM's phase transition and the thermal properties of the surrounding materials [58]. Using materials with higher thermal conductivity than $SiO_2$, or employing a PIN microheater, can improve the thermal time constant [68]. PCMs are ideal candidates for programmable PICs due to their superior thermal crosstalk immunity compared to other programming techniques [59]. The threshold-driven melt-quench process used for amorphization of PCMs requires temperatures above 500°C, making it highly resistant to thermal crosstalk. Consequently, any sub-threshold thermal interference does not impact the other densely packed on-chip components, ensuring crosstalk-free operation after programming [46]. Although crystallization has a lower temperature threshold (~200°C) and is more vulnerable to crosstalk, this issue can be effectively managed by initially crystallizing all devices and then selectively amorphizing the required ones [46,59]. This method maintains precise control and functionality of the PIC. maybe add one sentence about all the optical simulations are done under the assumption that the system is cooled down to room temperature, so no need to consider the thermal optical effect in the silicon. Note that all optical simulations (i.e., FDE mode analysis) were performed under the assumption that the system is cooled down to room temperature, so there is no need to consider the thermal optical effect in the silicon.

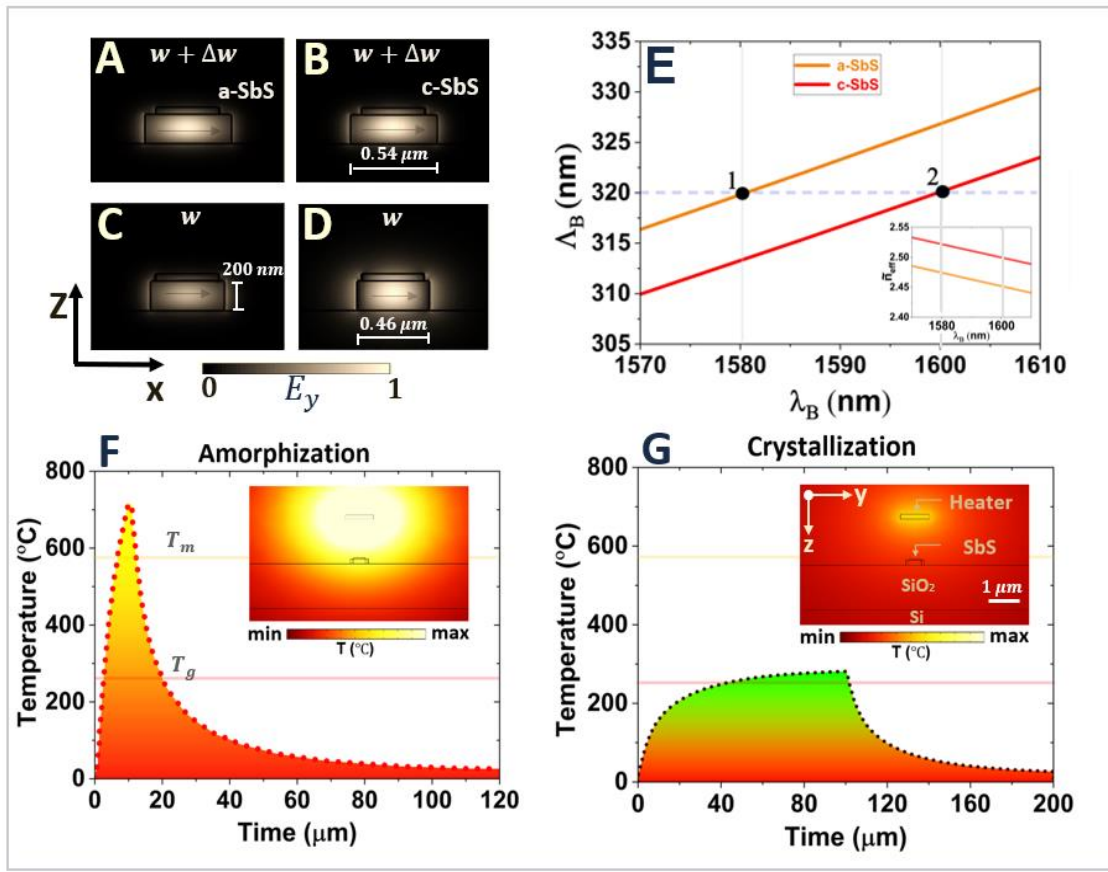

**Figure 2: Modes and Thermal Simulations of the proposed programmable filter.**
(A-C) Simulated mode profiles of the waveguide for different states of $Sb_2S_3$ with waveguide widths of 0.54 μm *(w+Δw)* and 0.46 μm *(w)*. The polarization of the electric field components is indicated by arrows in the embedded vector diagrams. (E) Tuning characteristics of the Bragg wavelength ($\lambda_B$) as a function of the Bragg period for both structural states of $Sb_2S_3$. The gray lines indicate the center of the Bragg wavelength, while the dashed gray line highlights the filter's grating period. A redshift in the Bragg wavelength from point 1 to point 2 is observed due to the phase transition of $Sb_2S_3$.

The inset illustrates the average effective index $\tilde{n}_{eff}$ of structure in both Sb₂S₃ states as a function of Bragg wavelengths. (F-G) Thermal simulation results showing the temperature profiles during the (F) amorphization and (G) crystallization processes of Sb₂S₃. The insets show the corresponding temperature distribution across the device, highlighting the areas affected by the heater. The graphs indicate the temperature evolution over time, with $T_m$ and $T_g$ representing the melting and glass transition temperatures, respectively. The quench time ($t_q$) is also displayed, showing the time it takes for the material to solidify after melting.

We evaluated the performance of the Sb₂S₃-based Bragg filter using simulations with ANSYS Lumerical's Eigenmode Expansion (EME) solver, focusing on key performance metrics: the full reflection spectrum in both states of Sb₂S₃, wavelength shift (*Δλ*), and bandwidth. The device cross-section was divided into two cell groups within the EME solver to improve computational efficiency: one for the wider waveguide region (w+Δw) and another for the narrower region (w). The grating period (*Λ*) was set to a fixed value (320nm), and the number of grating periods (*N*) was varied to investigate its impact on device performance. An initial periodicity of *N* = 300 was chosen, resulting in a unit cell (comprising two cell groups) being propagated 300 times and a total device length of approximately 300μm. This approach offers a comprehensive evaluation of filter performance while demanding less computational power than full-scale 3D Finite-Difference Time-Domain (FDTD) simulations. **Figure 3A** shows the reflectivity spectra for the filter in both amorphous (a-Sb₂S₃) and crystalline (c-Sb₂S₃) states, revealing a significant shift (*Δλ* = 20 nm) in the stopband wavelength for N = 300 periods. This stopband shift confirms the feasibility of using Sb₂S₃ phase transitions to tune the filter's spectral response. **Figure 3B** further examines the performance of the filter by presenting the reflectivity spectra for different grating periodicity (*N* = 50, 100, 200, 300, 500) in both structural states. The simulations reveal that increasing the grating length N results in a decrease in bandwidth and an increase in reflectivity and side lobes. The results in **Supplementary Figure 3** illustrate the impact of the number of grating periods on the bandwidth and peak reflectivity of the Bragg grating. As expected, increasing the number of periods enhances the peak reflectivity and reduces the bandwidth, which improves the filter's selectivity. However, this also introduces more pronounced side lobes. To mitigate side lobes in Bragg grating filters, several apodization methods are routinely employed [69–71].

To better understand light propagation within the filter, we utilized Ansys Lumerical's 3D Finite-Difference Time-Domain (FDTD) method to simulate the electric field distributions (**Figures 3C and 3D**). **Figure 3C** depicts the electric field distribution at the stopband wavelength. The observed intensity pattern confirms substantial reflectivity at this wavelength, signifying efficient light reflection by the filter. Conversely, **Figure 3D** depicts the electric field distribution at the passband wavelength, where light propagates through the waveguide with minimal reflection. It's noteworthy that these FDTD simulations were performed for $N$ = 30 periods due to the computationally intensive nature of 3D FDTD simulations for larger periodicities. Nevertheless, the results effectively demonstrate that the proposed $Sb_2S_3$ Bragg filter offers both efficient light filtering at the stopband and the potential for dynamic spectral tuning via the $Sb_2S_3$'s phase transition. **Supplementary Section 1** discusses the performance of the non-volatile programmable Bragg filter, focusing on the use of a PIN micro-heater to electrothermally switch $Sb_2S_3$.

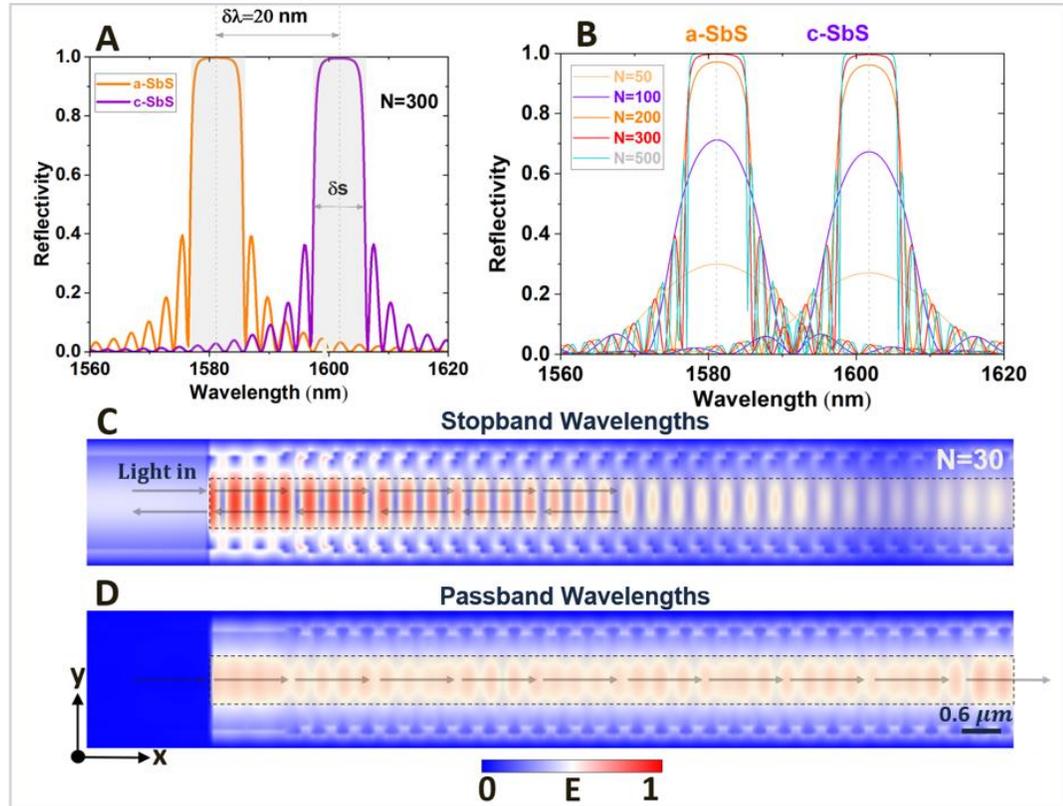

**Figure 3. Device performance.** (A) Reflectivity spectra for the Bragg filter in the amorphous (a-Sb$_2$S$_3$) and crystalline (c-Sb$_2$S$_3$) states, showing a tunable shift of $\Delta\lambda$ = 20 nm in the stopband wavelength for $N$ = 300 periods. (B) Reflectivity spectra for different numbers of grating periods (N = 50, 100, 200, 300, 500) in both amorphous and crystalline states of Sb$_2$S$_3$. (C-D) Simulated Electric field distribution along the propagation direction of the filter from right to left, where (C) stopband wavelengths, the high reflectivity, where light is reflected back, indicating that the wavelengths fall within the stopband of the Bragg grating, and (D) passband wavelengths, illustrates the transmission of light through the waveguide for wavelengths within the passband of the Bragg grating. The simulations are performed for $N$ = 30 periods, illustrating that even with a relatively small number of periods (N=30), the device can achieve significant filtering and transmission. Grey arrows indicate the propagation direction of the light, and dashed lines mark the Sb$_2$S$_3$ layer boundaries.

We also investigated the impact of varying the dimensions of the Sb$_2$S$_3$ layer over the stopband position. **Figures 4A and 4B** present central wavelength tuning maps for both the (a-Sb$_2$S$_3$) and (c-Sb$_2$S$_3$) states when varying the width and thickness of Sb$_2$S$_3$ dimensions. These maps reveal how the central wavelength of the stopband can be systematically shifted by controlling the thickness and width of the Sb$_2$S$_3$ layer. We further explored the quasi-continuous tuning capability of our device by leveraging the multilevel operation feature of $Sb_2S_3$. **Figure 4C** demonstrates the distinct reflectivity spectra of the filter. The difference in refractive index between the amorphous and crystalline states of $Sb_2S_3$ enables deterministic quasi-continuous, multilevel control over the reflectivity spectrum. This essentially translates to a 5-bit stable operation (32 levels) spanning from the fully amorphous state (step 1) to the fully crystalline state (step 32). Each phase corresponds to a specific step within the material's transition, resulting in fine-tuned adjustments to the reflectivity spectrum. It's important to note that the selection of a 5-bit operation reflects previously demonstrated experimental feasibility. **Figure 4D** highlights the resolution of this multi-level control, showing the 32 distinct levels possible during the Sb2S3 phase transition. Each level corresponds to a specific stopband central wavelength in the reflectivity spectrum. As $Sb_2S_3$ transitions from amorphous to crystalline, the central wavelength consistently redshifts. The figure shows this shift with a total change of 20 nm and an impressive resolution of 0.625 nm per step.

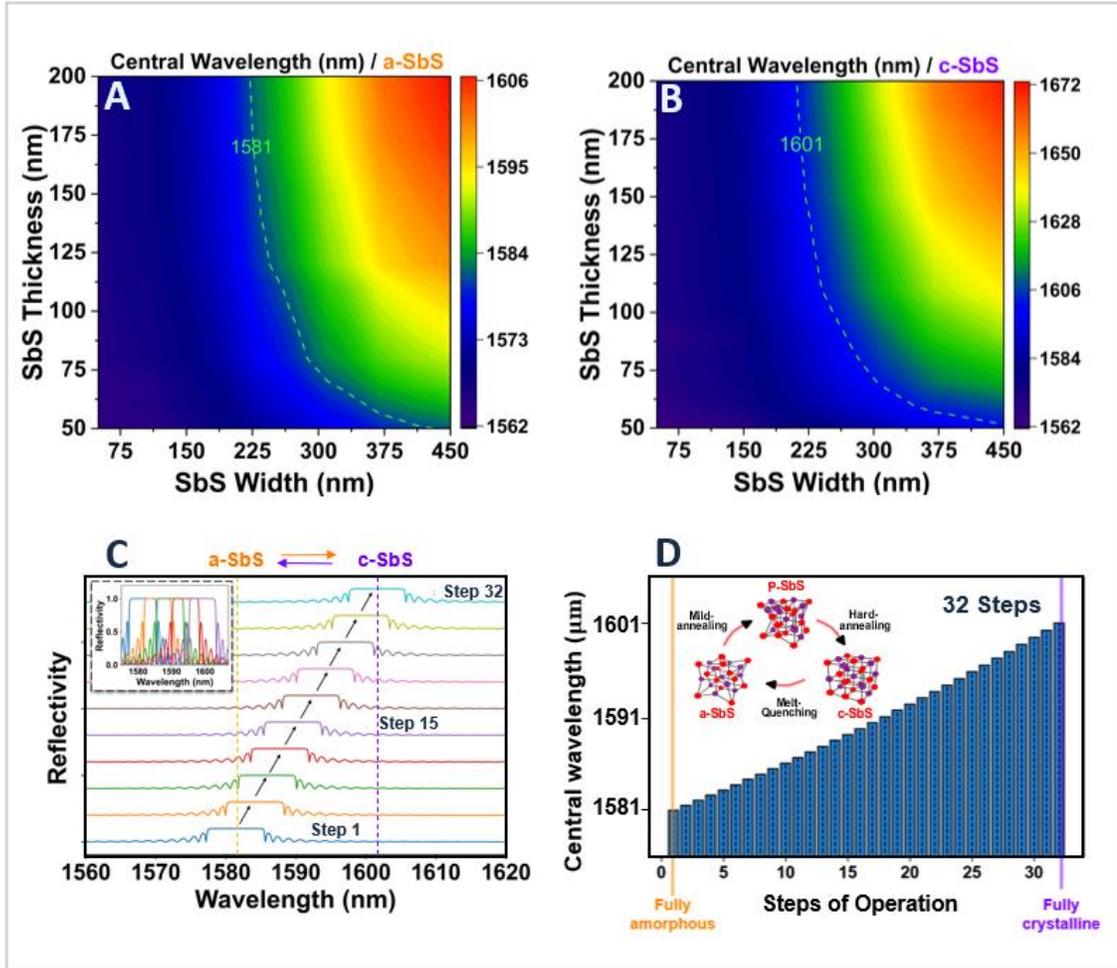

**Figure 4: Device Performance with Varying PCM Dimensions and Multilevel Operation.** (A-B) Central wavelength tuning maps for (A) amorphous (a-$Sb_2S_3$) and (B) crystalline (c-$Sb_2S_3$) states as functions of $Sb_2S_3$ thickness and width. The dashed lines indicate the nominal central wavelengths of the stopbands in both a-$Sb_2S_3$ and c-$Sb_2S_3$. (C) A stacked graph of the reflectivity spectra shows the distinct reflectivity spectrum for various steps of the distributed Bragg grating with $Sb_2S_3$ deposited on top. The ability to program $Sb_2S_3$ into 32 different states allows for fine-tuned adjustments of the reflectivity spectrum where each state corresponds to a specific step in the phase transition, resulting in incremental shifts in the reflectivity peaks. Inset the side-by-side version of the stacked figure. (D) 5-bit programmable operation illustrating the 32 levels of resolution achievable as $Sb_2S_3$ transitions from amorphous to crystalline form. Each step corresponds to a different central wavelength in the reflectivity spectrum. As the PCM is programmed from amorphous to crystalline, the central wavelength experiences a consistent redshift.

## 4. Conclusions:

In conclusion, we have presented the first non-volatile, electrically programmable on-chip Bragg filter. This device integrates a nanoscale $Sb_2S_3$ layer atop a periodically patterned silicon waveguide, achieving dynamic spectral tuning via integrated microheaters. Programmable electrical pulses induce localized Joule heating, triggering $Sb_2S_3$ phase transitions (amorphous-to-crystalline and vice versa) and modifying the refractive index profile of the grating. Our design offers a significant advancement over traditional passive Bragg gratings and active volatile filters, providing electrically controlled, reconfigurable spectral responses in a non-volatile manner. The filter demonstrates exceptional performance, with a peak reflectivity exceeding 99% and a tuning range of $\Delta\lambda$ = 20 nm during the phase transitions of $Sb_2S_3$. Additionally, we demonstrate quasi-continuous spectral control by manipulating the $Sb_2S_3$ phase distribution. A key feature of our approach is its low power consumption and immunity to thermal crosstalk, attributed to the non-volatile and threshold-switching nature of the PCM, addressing the significant issues of current tunable filters. Our programmable filter, with its high tuning range and fine spectral control, is poised to significantly impact a wide array of applications, including microwave photonics, optical signal processing, optical networks and adaptive multi-parameter sensing.